\documentclass[%
reprint,
 amsmath,amssymb,
 aps,
 longbibliography,
pra,
superscriptaddress
]{revtex4-1}

\usepackage{graphicx}
\usepackage{dcolumn}
\usepackage{bm}
\usepackage{hyperref}
\usepackage{braket}
\usepackage{xcolor}

\newcommand{\hjc}{H_\text{JC}}

\newcommand{\hcd}{H_\text{CD}}

\newcommand{\he}{H_\text{e}}
\newcommand{\ue}{U_\text{e}}
\newcommand{\me}{M_\text{e}}

\newcommand{\hcdp}{H_\text{p}}
\newcommand{\sz}[1]{\sigma_z^{(#1)}}
\newcommand{\splus}[1]{\sigma_+^{(#1)}}
\newcommand{\smin}[1]{\sigma_-^{(#1)}}
\newcommand{\ud}{U_\text{d}}
\newcommand{\If}{\mathbb{I}_f}
\newcommand{\av}{\langle \If(\bm{\epsilon})\rangle_\epsilon}

\begin{document}

\title{Accelerating adiabatic protocols for entangling two qubits in circuit QED}

\author{F. Petiziol}
\affiliation{%
 Dipartimento di Scienze Matematiche, Fisiche e Informatiche, Universit\`a di Parma, Parco Area delle Scienze 7/A, 43124 Parma, Italy
}
\affiliation{%
INFN, Sezione di Milano Bicocca, Gruppo Collegato di Parma, Parco Area delle Scienze 7/A, 43124 Parma, Italy
}
\author{B. Dive}%
\affiliation{%
Department of Physics, Imperial College London, SW7 2AZ London, UK
}
\affiliation{%
Institute of Quantum Optics and Quantum Information, Boltzmanngasse 3, 1090 Vienna, Austria}
\author{S. Carretta}
\affiliation{%
 Dipartimento di Scienze Matematiche, Fisiche e Informatiche, Universit\`a di Parma, Parco Area delle Scienze 7/A, 43124 Parma, Italy
}
\author{R. Mannella}
\affiliation{
Dipartimento di Fisica, Universit\`a di Pisa,  Largo Bruno Pontecorvo 3, 56127 Pisa, Italy
}
\author{F. Mintert} 
\affiliation{%
Department of Physics, Imperial College London, SW7 2AZ London, United Kingdom
}
\author{S. Wimberger}
\affiliation{%
 Dipartimento di Scienze Matematiche, Fisiche e Informatiche, Universit\`a di Parma, Parco Area delle Scienze 7/A, 43124 Parma, Italy
}
\affiliation{%
INFN, Sezione di Milano Bicocca, Gruppo Collegato di Parma, Parco Area delle Scienze 7/A, 43124 Parma, Italy
}
 \email{sandro.wimberger@unipr.it}

\date{\today}

\begin{abstract}
We introduce a method to speed up adiabatic protocols for creating entanglement between two qubits dispersively coupled to a transmission line, while keeping fidelities high and maintaining robustness to control errors. The method takes genuinely adiabatic sweeps, ranging from a simple Landau-Zener drive to boundary cancellation methods and local adiabatic drivings, and adds fast oscillations to speed up the protocol while canceling unwanted transitions. We compare our protocol with existing adiabatic methods in a state-of-the-art parameter range and show substantial gains. Numerical simulations emphasize that this strategy is efficient also beyond the rotating-wave approximation, and that the method is robust against random static biases in the control parameters and with respect to damping and decoherence effects.
\end{abstract}
\maketitle

\section{\label{sec:level1}Introduction}
The development of quantum technologies depends crucially on the capability of realizing many high-quality logical operations on a quantum system within the coherence times. There are thus two objectives to be realized simultaneously, time and fidelity, which are often the opposite sides of the same coin: Doing a protocol slowly often reduces the associated gate errors, but limits the number of operations that can be performed within the natural lifetime of the system. In addition to this, there is the constraint that a quantum control problem is always subject to limitations on the control resources, that is, on the knobs which one can work with in order to influence the system's behavior.

If the time and control restrictions are not severe, adiabatic methods are a very interesting option due to their intrinsic robustness. They require a slow driving of the system Hamiltonian in time such that the system, as predicted by the adiabatic theorem of quantum mechanics \cite{messiah1961qm}, always remains in an instantaneous eigenvector. However, adiabatic methods may be too slow to be a viable strategy.

On a different side of the quantum control spectrum, the most used protocols exploit sequences of evolutions under time-independent effective Hamiltonians. These are typically faster and require less control resources, but they are often unstable with respect to biases in control and system parameters.

In this work we contribute to fill the gap between these two scenarios, by discussing a control protocol producing high fidelities in intermediate-range timescales, while still partially inheriting the robustness of adiabatic strategies and without requiring the addition of new ``control knobs.'' The protocol is contextualized in the problem of producing entanglement between two qubits in a standard circuit QED (cQED) architecture, namely constituted of two superconducting qubits dispersively coupled to a planar transmission line resonator \cite{Blais2007,GU20171}.  

We first compare different finite-time adiabatic sweeps, ranging from the standard linear Landau-Zener \cite{nori} to boundary cancellation methods (BCMs) \cite{Lidar2009} and local adiabatic drivings (LADs) \cite{Roland2002, Rezakhani2009}. These methods provide satisfactory final fidelities, for the experimental parameters considered, only for total durations exceeding 1 $\mu$s, that is, for a total time which is rather long for performing a single operation in the experimental platform considered. For this reason, we then study how the protocol can be sped up by allowing the presence of a degree of nonadiabaticity in the process which is dynamically counteracted by an additional correcting control Hamiltonian. The latter can be absorbed into the original Hamiltonian by adding quickly oscillations in the initial parameters. This strategy was introduced in Ref. \cite{petiziol2018} and is built upon the theory of counterdiabatic (CD) or transitionless quantum driving \cite{rice1,berry1}, but, contrary to this one, it does not require the introduction of new Hamiltonian components. It is realizable instead, in the setup under analysis, by allowing the qubit-bus couplings to oscillate quickly. For brevity, we will refer to this method as effective counterdiabatic (eCD) method. Similar ideas in the use of fast oscillating controls for accelerating adiabatic dynamics were also recently pursued in Refs. \cite{Boyers2018,Bukov2018}, based on Floquet-engineering techniques \cite{Bukov2015}.

We show that, by adding eCD corrections to standard adiabatic drivings, high fidelities can be attained much more quickly. In particular, by integrating the eCD scheme with a LAD, timescales of practical interest can be achieved while remaining within state-of-the-art values of the system parameters. At the same time, robustness against systematic errors of the latter protocol is proven by showing that small random imperfections in the control parameters only weakly affect the final fidelity. High fidelities are maintained also in the presence of relaxation and decoherence of the two qubits and photon leakage from the resonator, as shown by numerical solution of a proper Lindblad-type master equation.

The physical system under analysis is introduced in Sec. \ref{sec:model}, where also the basics of the adiabatic control protocol are presented, and different choices for the adiabatic sweep function are discussed. In Sec. \ref{sec:cd}, the CD method is reviewed before describing the eCD method.
The method is then applied to speed up the adiabatic process and the results are discussed in Sec \ref{sec:results}, where it is shown that best performance is achieved by combining eCD corrections with a local adiabatic driving. The robustness of these results against imperfections in the control parameters, and against dissipation and decoherence effects is analyzed in Sec. \ref{sec:stability}, followed by the conclusion in Sec. \ref{sec:conclusion}.

\section{The model}\label{sec:model}
\subsection{General setup}
The general model we will consider is that of two (artificial) atoms whose interaction is mediated by a single bosonic mode acting as quantum bus. The Hamiltonian $\mathcal H$ thus includes two anharmonic ladders representing the atoms, described by field operators $b_{k}$ and $b_k^{\dagger}$ with $k=1,2$, and a single-mode harmonic oscillator described by field operators $a$ and $a^{\dagger}$ \cite{Blais2007}: In units of $\hbar$, it reads
\begin{align}
\mathcal H= & \sum_{k=1,2} \left[ \omega_k b^{\dagger}_k b_k +\frac{\alpha_k}{2} b_k^{\dagger} b_k^{\dagger} b_k b_k \right] + \omega_r a^{\dagger} a\nonumber\\
 & + \sum_{k=1,2} g_k(b_k+b_k^{\dagger})(a + a^{\dagger}). \label{eq:hAmin2009it}
\end{align}
The parameter $\omega_k $ denotes the transition frequency of the two lowest levels of atom $k$, $\omega_r$ is the oscillator transition frequency, $\alpha_k$ denotes the anharmonicity of the atoms' energy levels, and $g_k$ is the coupling between atom $k$ and the resonator.
Performing a two-level approximation for the two atoms and assuming the rotating-wave approximation (RWA), the Hamiltonian of Eq. \eqref{eq:hAmin2009it} can be rewritten in Jaynes-Cummings form \cite{mandel_wolf_1995,Wendin2017,GU20171},
\begin{equation}\label{eq:hjc}
\hjc = \omega_r a^{\dagger} a + \sum_{k=1}^2 \frac{\omega_k}{2} \sz{k} + \sum_{k=1}^2 g_k [\splus{k} a + \smin{k} a^{\dagger}],
\end{equation}
where $\{\sigma_x^{(k)},\sigma_y^{(k)},\sigma_z^{(k)}\}$ are the Pauli matrices acting on the Hilbert space of qubit $k$ and $\sigma_\pm^{(k)}=[\sigma_x^{(k)}\pm i \sigma_y^{(k)}]/2$ are the corresponding raising and lowering operators.
This Hamiltonian can physically describe, for instance, atoms in a cavity \cite{Haroche2006}, atoms \cite{Haffner2008} and trapped ions \cite{ReviewRydberg} under external laser fields, or superconducting qubits coupled to transmission line resonators \cite{Blais2007,Majer2007,Devoret2013,Wendin2017}. The latter is the specific framework which we will take into account in this work. In fact, these platforms have proved to be extremely promising candidates for building quantum technology, especially because of their scalability as solid-state devices and their relatively long coherence times \cite{Devoret2013,Clarke2008,Schoelkopf2008,Gambetta2017}. They have been a warhorse for the experimental investigation of many complex quantum phenomena, such as basic quantum algorithms \cite{DiCarlo2009} and quantum error correction \cite{Reed2012,DiCarlo2010}, quantum teleportation \cite{Steffen2013}, quantum simulation \cite{Langford2017}, multi-qubit entanglement \cite{Barends2014, DiCarlo2010} and Jaynes-Cummings physics \cite{Fink2008}. 

We will assume tunability both of the internal transition frequency of the qubits, $\omega_k=\omega_k(t)$, and of the qubit-resonator couplings, $g_k=g_k(t)$. The former is standard for typical superconducting qubits, such as the transmon \cite{Koch2007}. This is a Cooper-pair box charge qubit \cite{Clarke2008} shunted by an additional high capacity which allows large ratios $E_J/E_C$ between Josephson energy $E_J$ and charging energy $E_C$, as compared to a standard Copper-pair box. This makes it rather insensitive with respect to low-frequency charge noise. As a result, it exhibits long coherence times with respect to standard charge, flux and phase qubits \cite{Koch2007}. The transmon can be loosely though of as a LC electric circuit with the inductance replaced by a Josephson junction which induces the anharmonicity of the energy levels. 

Tunability of the coupling can be achieved by replacing, for instance, the standard transmon with a so-called tunable coupling qubit (TCQ) proposed in Refs. \cite{Gambetta2011,Srinivasan2011}.
The TCQ is constituted of three superconducting islands connected by Superconducting Quantum Interference Devices (SQUIDs). As suggested in Ref. \cite{Srinivasan2011}, it can be thought of as a system composed of two transmonic qubits capacitively coupled. 
By changing the energies of the islands, which is done by varying the energy and external magnetic flux applied to the SQUIDs, one can change the dipole moments of the two transmons. These in turn can combine in a parallel or antiparallel way, resulting globally in a dipole or quadrupole moment, respectively. In the parallel case, the TCQ effectively couples strongly to the resonator, while in the antiparallel case it does not. As a result, tuning between the two configurations allows one to tune the TCQ-resonator coupling.

\subsection{Specific framework} \label{sec:specfram}
\begin{figure}
\includegraphics[width=0.9\linewidth]{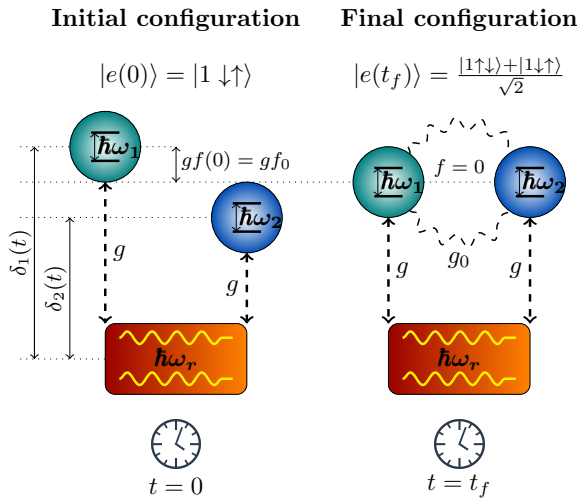}
\caption{Sketch of the general setup. The two qubits, having transition frequencies $\omega_1$ and $\omega_2$, are coupled with strength $ g$ to a resonator of transition frequency $\omega_r$, being detuned from it by $\delta_1$ and $\delta_2$. The two qubits are initially, at time $t=0$, detuned in frequency from each other by $\omega_1(0)-\omega_2(0) = 2 g f_0$ (left). They are then slowly driven into resonance, getting such at time $t=t_f$ (right). Initially, if $\omega_1>\omega_2$ the (instantaneous) first excited state is $\ket{e(0)}=\ket{1 \downarrow \uparrow}$, while the final (instantaneous) excited state at resonance is $(\ket{1\uparrow \downarrow}+\ket{1\downarrow \uparrow})/\sqrt{2}$. The instantaneous qubit detuning is described by the sweep function $f(t)$ which drives the qubits slowly into resonance with each other, so that $f(0)=f_0$ and $f(t_f)=0$.}
\label{fig:sketch}
\end{figure}
Starting from the Hamiltonian of Eq. \eqref{eq:hjc}, let us restrict the discussion to the two-excitation manifold: The Hamiltonian is then a four-by-four matrix. Let us move to a reference frame rotating at the resonator frequency $\omega_r$ and introduce the qubit-resonator detunings $\delta_k=\omega_k-\omega_r$. In the dressed basis $\{\ket{0\uparrow \uparrow}, \ket{1\uparrow\downarrow}, \ket{1\downarrow\uparrow}, \ket{2\downarrow\downarrow} \}$, where the first quantum number indicates the number of field excitations while the others refer to the two qubits bare states, the Hamiltonian explicitly reads
\begin{equation}\label{eq:ham}
H= \begin{pmatrix}
\frac{\delta_1 + \delta_2}{2} & g_2 & g_1 & 0 \\
g_2 & \frac{\delta_1-\delta_2}{2} & 0 & \sqrt{2} g_1 \\
g_1 & 0 & \frac{\delta_2-\delta_1}{2} & \sqrt{2}g_2 \\
0 & \sqrt{2} g_1 & \sqrt{2} g_2 & - \frac{\delta_1+\delta_2}{2}
\end{pmatrix}.
\end{equation}
In the dispersive regime $g_k/\delta_k \ll 1$, an avoided crossing is formed when the two qubits are brought into resonance far from the resonator's transition frequency \cite{Blais2004,Blais2007,Majer2007}. This effect shows the evidence of a bus-mediated coupling between the two qubits, produced by the exchange of a virtual photon. Such a process is well understood by second order perturbation theory in $g_k/\delta_k$ \cite{Blais2007}. As shown in Appendix \ref{appendix:AC}, by applying the transformation $U = \exp\left[\sum_k \frac{g_k}{\delta_k}(a \splus{k}-a^{\dagger} \smin{k})\right]$ to \eqref{eq:hjc}, the Hamiltonian $U H U^{\dagger}$ to second order exhibits a qubit interaction term $\propto \splus{1}\smin{2}+\smin{1}\splus{2}$ \cite{Blais2004,Majer2007}. The latter is responsible for the formation of the anticrossing, and can produce entanglement between the qubits. For example, it can be exploited to realize a $\sqrt{i\text{SWAP}}$ gate \cite{Majer2007}. We denote the width of the anticrossing by $2 g_0$, with $g_0$ being positive. 

Here, our purpose is to take advantage of the above described avoided crossing for the adiabatic preparation of an entangled Bell state.
Rather than kicking the qubits fast into resonance with each other, and then letting them evolve freely under the time-independent Hamiltonian, the idea is to escort them slowly, so that the system always remains in an instantaneous eigenstate of the evolving Hamiltonian as predicted by the adiabatic theorem. If this is the case, the system, initially assumed to be in the first excited state, which is $\sim \ket{1 \downarrow \uparrow}$ if $\omega_1>\omega_2$, will end up at resonance in the entangled (instantaneous eigen)state  $\sim(\ket{1\uparrow \downarrow} + \ket{1\downarrow \uparrow})/\sqrt{2}$. Alternatively, starting from the second excited $\sim \ket{1 \uparrow \downarrow}$ would produce the singlet state $(\ket{1\uparrow \downarrow} - \ket{1\downarrow \uparrow})/\sqrt{2}$. Entanglement preparation via adiabatic methods in cQED was also studied in different frameworks and architectures; see, for instance, Refs. \cite{DiCarlo2009, YuChen2014, Martinis2014, Xu2016}.

For achieving satisfactory final fidelities, slowly means typically too slow though with respect to reasonable coherence times. For this reason, a shortcut-to-adiabaticity method first introduced in Ref. \cite{petiziol2018} will be used which tracks to the true adiabatic path closely without requiring the introduction of the new terms in the Hamiltonian \eqref{eq:ham}. It requires, in exchange for that, fast oscillation of the available terms. 

The setup is the following. Let the protocol have a total duration $t_f$. Let the two qubit-bus couplings have the same value, $g_1=g_2 \equiv g$, and let all quantities be rescaled by $g$. The whole protocol proposed here can be formulated also for the case in which $g_1\ne g_2$ by choosing a different final configuration of the system. Specifically, one should determine the configuration such that the final instantaneous eigenvector is the desired Bell state, which for $g_1\ne g_2$ occurs when the qubits are not exactly in resonance with each other. We introduce a sweep function 
\begin{equation}\label{eq:sweep0}
f(t) = (\delta_1-\delta_2)/2g
\end{equation}
which drives the (rescaled) qubit detuning $(\delta_1-\delta_2)/g$ from an initial (positive) value $2 f_0$ to zero. When we rescale the physical time like $s=t/t_f$ and introduce $\tau=g t_f$ and $\delta_{g} =  [\delta_1(t)+\delta_2(t)]/2g$, the dimensionless Schr\"odinger equation for the evolution operator reads
$$ i \partial_s U(s) = \tau H(s) U(s) $$
with
\begin{equation}\label{eq:syst}
H(s) =  \begin{pmatrix}
 \delta_{g}  & 1 & 1 & 0 \\
1 & f(s) & 0 & \sqrt{2}  \\
1 & 0 &-f(s)& \sqrt{2} \\
0 & \sqrt{2} & \sqrt{2} & -  \delta_{g}
\end{pmatrix}.
\end{equation}
Using the matrices $\{D_0,D_1,C_1,C_2\}$ introduced in Appendix \ref{sec:matrices}, this Hamiltonian can be expressed more compactly like
$$H(s)= \delta_{g} D_0 + f(s) D_1 + C_1 + C_2. $$
A sketch of the whole setup with the relevant quantities is represented in Fig. \ref{fig:sketch}.
An experimental setup very similar to the one considered here was used in Ref. \cite{Quintana2013} for studying Landau-Zener-St\"uckelberg effects at the avoided crossing. In the following, the parameters which will be used for all the simulations are taken to be $\omega_r/2\pi = 8.2$ GHz, $g/2\pi=50$ MHz, and the qubit transition frequencies $\omega_1/2\pi$, $\omega_2/2\pi$ sweep from $6.01$ and $5.99$ GHz, respectively, to $6$ GHz at resonance producing an avoided crossing of width $2g_0 \sim14$ MHz. These values are chosen among typical parameter ranges used for cQED systems \cite{Majer2007,Schoelkopf2008,DiCarlo2009}. Translated into the dimensional quantities appearing in Eq. \eqref{eq:syst}, they give $\delta_g = -44$, $f_0 = 0.2$.

\subsection{Sweep functions}\label{sec:sweeps}
\begin{figure}
\includegraphics[width=\linewidth]{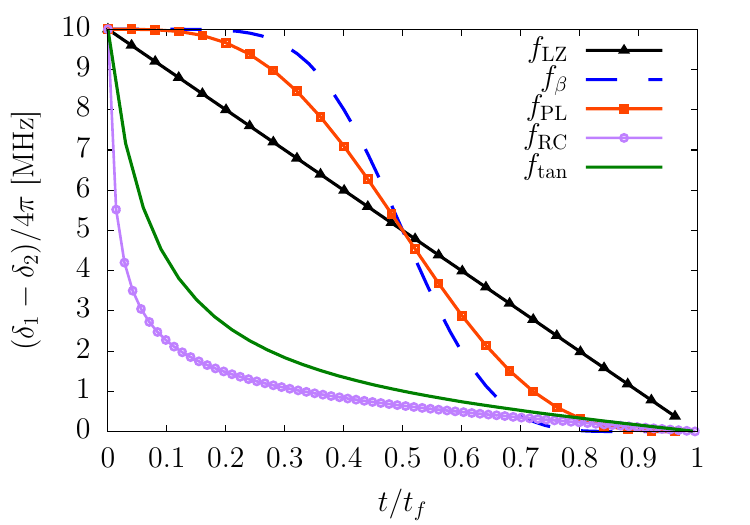}
\caption{(Rescaled-)time profiles of the adiabatic sweep functions considered in Sec. \ref{sec:sweeps}. Going beyond a simple linear Landau-Zener driving $f_\text{LZ}$, the functions $f_\beta$ and $f_\text{PL}$ represent boundary cancellation methods, while $f_\text{RC}$ and $f_\text{tan}$ are local adiabatic drivings.}
\label{fig:sweep}
\end{figure}

Here we discuss different choices for the finite-time adiabatic sweep function $f(s)$ of Eq. \eqref{eq:sweep0} which brings the two qubits into resonance starting from an initial rescaled gap $(\delta_2-\delta_1)/g=2f_0$, so that $f(0)=f_0$ and $f(1)=0$. We compare their performance in terms of final fidelity with respect to the total duration of the sweep. Specifically, the possibilities that are taken into account are the following, which are represented in Fig. \ref{fig:sweep}:
\begin{enumerate}
   \item[(1)] \emph{Linear (Landau-Zener-like):} 
    $$ f_\text{LZ}(s) = f_0[1-s] .$$
    \item[(2)] \label{poly} \emph{Polynomial}, with three derivatives vanishing at the beginning and at the end of the sweep.
 	   \begin{equation}\label{eq:poly}
 	    f_\text{PL}(s) = f_0[1-35 s^4+84 s^5-70 s^6+20 s^7]. 
 	    \end{equation}
    Polynomials with higher and lower numbers of zero derivatives were also considered. The lower number case performed worse, while there was negligible improvement in the higher number case.
    \item[(3)] \label{regbeta} \emph{Regularized Beta function:}
    $$f_\beta(s) = f_0[1-\Theta_k(s)]$$
    with
    $$ \Theta_k(x) = \frac{B_s(1+k,1+k)}{B_1(1+k,1+k)}, $$
    where $B_s(a,b) = \int_0^s y^{a-1}(1-y)^{b-1} dy$, with $\text{Re}(a)>0$, $\text{Re}(b)>0$, $|s|\le 1$ is the incomplete Beta function \cite{weber}.
    This function was proposed in Ref. \cite{Rezakhani2010} as an adiabatic sweep function and has $k$ derivatives vanishing at the boundary. We will show the results obtained with $k=8$, since larger $k$ did not result to improve the performance in our cases. 
     \item[(4)] \label{rcf}\emph{Roland-Cerf (RC):}
    \begin{equation}\label{eq:rc}
      f _\text{RC}(s) = \frac{g_0 f_0(1-s)}{g \sqrt{(g_0/g)^2+ f_0^2 s(2-s)}},
    \end{equation}
     where $2 g_0$, introduced in Sec. \ref{sec:specfram}, is the minimum gap between the first and second excited states in the final configuration -- that is for $f(s)=0$. The value of $g_0$ can be approximately determined by perturbative arguments, see Appendix \ref{appendix:AC} and Ref. \cite{Blais2004}. Here the value is extrapolated from numerical diagonalization.
    More details on the derivation of the functional dependence in Eq. \eqref{eq:rc} are in Appendix \ref{sec:lad}.
 \item[(5)] \label{funtan} \emph{Tangent}:
 \begin{equation}\label{eq:qab}
 f(s) = \frac{g_0}{g} \tan [\alpha (1-s)],
 \end{equation}
 with $\alpha = \arctan(g f_0/g_0)$ and $g_0$ is as in point \ref{rcf}. The derivation of this function in connection to local adiabaticity is given in Appendix \ref{sec:lad}.
\end{enumerate}
\begin{figure}
\includegraphics[width=\linewidth]{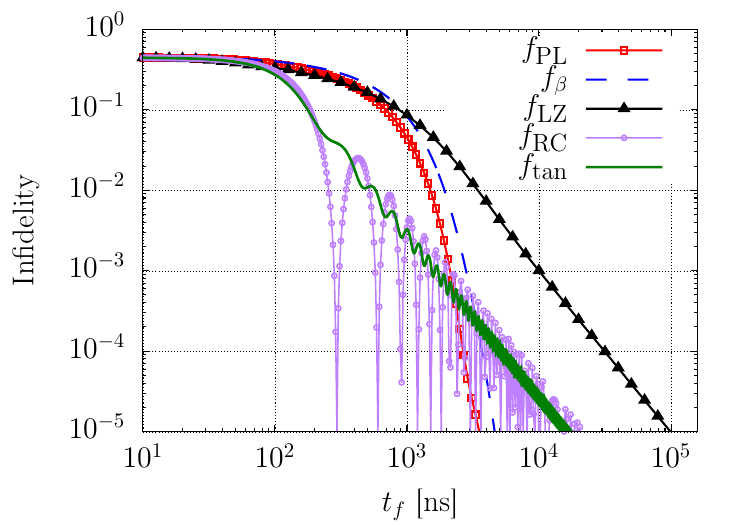}
\caption{Infidelity [see Eq. \eqref{eq:inf}] as a function of the total duration of the protocol for different adiabatic sweep functions. The use of boundary cancellation methods $f_{PL}$ and $f_\beta$ (red squares and blue dashed lines, respectively) strongly increases the performance with respect to a simple Landau-Zener sweep $f_\text{LZ}$ (black triangles line). For faster times, the local adiabatic drivings $f_\text{RC}$ and $f_\text{tan}$ (purple circles and green solid lines, respectively) give further improvement.}
\label{fig:adiabatic}
\end{figure}
	Cases \ref{poly} and \ref{regbeta} belong to the class of so-called boundary cancellation methods (BCMs) \cite{Lidar2009}. The latter predict that, once a duration $t_f$ sufficient to satisfy the global adiabatic condition (see Appendix \ref{sec:lad}) is chosen, greater (arbitrary, in principle) precision can be achieved by setting to zero an always larger number of derivatives at the beginning and at the end \cite{Rezakhani2009,Lidar2009}. Cases \ref{rcf} and \ref{funtan} represent local adiabatic drivings (LADs) instead, in which the instantaneous velocity of the evolution is adapted such as to satisfy a local adiabatic condition \cite{Roland2002,Rezakhani2009} (see again Appendix \ref{sec:lad}). The RC method was proven to be optimal for a one-dimensional (1D) quantum search problem with respect to a time-optimal adiabatic variational principle in Ref.  \cite{Rezakhani2009}. A RC function for a Landau-Zener avoided crossing problem was also studied, both theoretically and experimentally, in Refs. \cite{bason1,malossi1}.

The results produced by all the sweep functions considered here can be compared in Fig. \ref{fig:adiabatic}, where the final infidelity, defined as
\begin{equation}\label{eq:inf}
\If = 1-\lvert \braket{ \psi(t_f) \lvert e(t_f)} \lvert^2. 
\end{equation}
with $\ket{\psi(t_f)}$ the final system state and $\ket{e(t_f)}$ the final first excited state that we wish to prepare, is shown as a function of the total duration of the protocol. Figure \ref{fig:adiabatic} is produced by solving numerically the Schr\"odinger equation for different sweep functions and different total time $t_f$. The use of the functions $f_\text{PL}$ and $f_\beta$ (red squares and blue dashed line, respectively) produce almost equivalent performance, and they prove to dramatically increase the precision with respect to a simple Landau-Zener (LZ) sweep at fixed protocol duration. For example, an infidelity of $10^{-3}$, requiring $10$ $\mu$s if a LZ sweep is used, can be attained in $\sim 2-3$ $\mu$s using $f_{PL}$ or $f_\beta$. Further improvement is obtained for shorter times using the RC or tangent sweeps $f_\text{RC}$ and $f_\text{tan}$ (green-solid and purple-circles, respectively). For instance, an infidelity of $10^{-2}$ can be attained in $\sim 0.5$ $\mu$s , while the same precision would require $\sim 1-2$ $\mu$s and $\sim 2.5$ $\mu$s using BCMs and a LZ sweep, respectively. Therefore, if one can accept timescales around 1 $\mu s$, global adiabatic drivings are the best choice, while LADs are to be used for ulteriorly speeding up the process. One can see that the RC sweep produces fidelities highly fluctuating with respect to the final duration. In principle, this effect could be exploited by selecting a duration close to a local minimum of the infidelity, but the high sensitivity suggests that this would be very hard in practice, and the protocol would still be quite unstable with respect to control imperfections.

\section{Counterdiabatic control fields}\label{sec:cd}
\begin{figure}
\includegraphics[width=\linewidth]{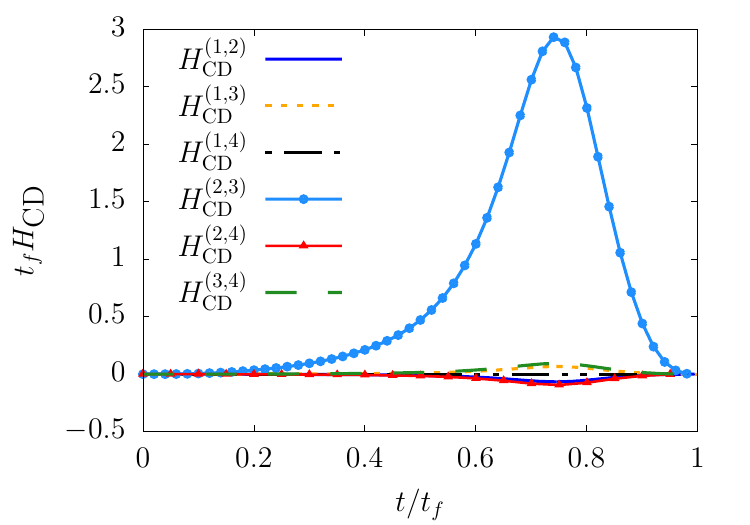}
\caption{Imaginary part of the matrix components of $\hcd$ (real parts are zero), see Eq. \eqref{eq:imhcd}. The (2,3) entry is the dominant one. Taking also into account that this is the nonadiabatic coupling between the two levels realizing the minimal energy gap (in the final configuration), $H_\text{CD}^{(2,3)}$ is by far the dominant contributor to nonadiabatic transitions.}
\label{fig:CDfield}
\end{figure}

In this section, we discuss the use of shortcuts to adiabaticity \cite{TORRONTEGUI2013117} for speeding up the desired adiabatic state transfer and entanglement production. The first case which we treat is that of the CD quantum driving \cite{rice1,berry1} theory. This provides a nonadiabatic method which guarantees unit fidelity in arbitrary time, but at the cost of introducing many new time-dependent control components and, for very fast times, strong correcting fields \cite{Theisen2017,petiziol2018,delcampo2013}. Implementation of the full protocol turns out to be too demanding in the experimental platform considered here, so we will address the case in which only a partial, dominant correction is taken into account. Since this is in turn hard to realize in practice, we will make use of the recently proposed method of effective counterdiabatic driving \cite{petiziol2018}. This allows one to produce the needed correction terms in an effective manner by introducing fast oscillations in the already available parameters only. 

\subsection{Exact counterdiabatic corrections}
Let us derive the CD correcting field for the setup described in Sec. \ref{sec:model}: This will be the starting ingredient for determining the final eCD Hamiltonian.

Let the original Hamiltonian of the system be
$$ H(t) = \sum_{n=1}^N E_n(t) \ket{n_t}\bra{n_t}.$$
The theory of counterdiabatic (CD) fields \cite{rice1,berry1} provides a recipe for finding a Hamiltonian $\hcd(t)$ which drives the eigenstates of the original Hamiltonian $H(t)$ with no transitions in a desired time.

Given a matrix $\ud(t)$ which diagonalizes $H(t)$ at time $t$,
$$\ud^{\dagger}(t) H(t) \ud(t) = \text{diag}\{E_1(t),\dots,E_N(t) \},$$
the CD field has the form
$$ \hcd(t) = i \partial_t \ud(t) \ud^{\dagger}(t) .$$
A more convenient expression, in terms of the instantaneous spectral properties of $H(t)$, is \cite{berry1}
	\begin{equation}\label{eq:hcdt}
		\hcd = i \sum_{m\ne n}\sum_{n=1}^N \frac{\ket{m_t}\bra{m_t} \partial_t H \ket{n_t}\bra{n_t}}{E_n(t)-E_m(t)}. 
	\end{equation}
Exact CD driving has been investigated experimentally, for instance, in Ref. \cite{bason1} and, using superconducting qubits, in Ref. \cite{paraonau2018,Danilin2018,paraonau2019}.
	
Let us now focus on the system of Eq. \eqref{eq:syst}. For the rest on this subsection, the reference adiabatic sweep we will work with is $f_\text{PL}$ as defined in Eq. \eqref{eq:poly}. 
The CD field cannot be computed analytically, since the Hamiltonian matrix cannot be generally diagonalized by exact methods. Therefore, inspection of its structure is done numerically. 
From general results on CD fields \cite{petiziol2018}, one already knows that $\hcd$ will have only imaginary matrix entries, since $H(t)$ is real. Therefore, its components are hard to be realized within the specific experimental setup considered here. New control elements should be introduced, or one should radically change the whole architecture. This is the reason why we will treat the problem via the eCD method in the next section. The (imaginary part of the) matrix elements
\begin{equation}\label{eq:imhcd}
\hcd^{(i,j)}(s) = \text{Im} \bra{i} \hcd(s) \ket{j}
\end{equation}
are plotted in Fig. \ref{fig:CDfield}.
One immediately sees that the largely dominant component is a coupling between the bare states $\ket{1\uparrow\downarrow}$ and $\ket{1 \downarrow \uparrow}$, $\hcd^{(2,3)}$. Taking also into account that the corresponding two levels are those with minimal gap, it is easy to recognize that this component gives the by far dominant contribution to nonadiabatic errors. Such a component is enhanced, while the others are further suppressed, if the detuning $\delta_{1,2}$  between qubits and the resonator increases.

In the following, we will thus concentrate on the effect of $\hcd^{(2,3)}$. In particular, this will be compensated for via the eCD method, while the others will be neglected. This corresponds to taking a two-level approximation focusing on the subspace spanned by $\ket{1\uparrow \downarrow}$ and $\ket{1 \downarrow \uparrow}$, while neglecting transitions outside this space. However, we still work with the numerical values obtained with the full matrix, rather than using an adiabatically reduced two-level matrix. 
For the moment, let us define the ``partially'' correcting CD field $\hcdp$, defined as
\begin{equation}\label{eq:hcdp}
 \hcdp(s) =  \frac{\hcd^{(2,3)}}{2}[\sigma_x^{(1)}\sigma_y^{(2)}-\sigma_y^{(1)}\sigma_x^{(2)}],
 \end{equation}
which equals $\hcd$ with $\hcd^{(i,j)}=0$ for $(i,j)\ne(2,3)$ or $(3,2)$.

The evolution of the system under the action of $H(s)+\hcdp(s)$ gives at least order-$10^{-5}$ infidelity for $t_f>1$ ns, and still improves as the total duration increases. The deviation of the infidelity from zero quantifies the error due to neglecting all nonadiabatic couplings apart from the one connecting first and second instantaneous excited states. The CD method would thus give great results if one was able to implement at least the $H_\text{CD}^{(2,3)}$ matrix elements. Still, the total duration affects the strength of the corrective field: The faster the protocol, the stronger $\hcd$ is, since $\hcd$ scales like $1/t_f$ \cite{berry1}. A measure of the magnitude of the matrix elements of $\hcd$ for our parameters can be estimated from Fig. \ref{fig:CDfield}: It is the value represented in the figure divided by $t_f$, in units of frequency. For example, achieving a fidelity greater than $99.999\%$ in 50 ns would require a $\hcd^{(2,3)}$ component with maximal amplitude around 10 MHz. In the architecture under consideration thought, the necessary strong coupling matrix elements are not available, and for this reason we introduce the eCD method in the next section. 

\subsection{eCD corrections}\label{sec:ecd}

The eCD method which will be used now has been introduced in Ref. \cite{petiziol2018}, for realizing a CD Hamiltonian in an approximate manner, by working only with the initially available control Hamiltonians. This kind of problem was also addressed in Ref. \cite{opatrny201}, using different methodologies. A summary of the general procedure for constructing the correcting Hamiltonian is the following:
\begin{enumerate}
\item[(1)] Write an ansatz Hamiltonian $\he$ constituted of the available control Hamiltonians $\{D_0,D_1,C_0,C_1\}$ given in Appendix \ref{sec:matrices}, controlled by means of control functions $\{d_0(t),d_1(t),c_0(t),c_1(t)\}$, respectively,
$$ \he = d_0(t) D_0 + d_1(t) D_1 + c_0(t) C_0 + c_1(t) C_1. $$
\item[(2)] Choose periodic control functions, of period $\omega=2 \pi/T$, written in the form of a truncated Fourier expansion
	\begin{equation}\label{eq:truncfour}
		 \sum_{k} \left[\mathcal A_{k} \sin(k \omega t)+\mathcal B_k \cos(k \omega t) \right] ;
	 \end{equation}
\end{enumerate}
\begin{enumerate}	 
\item [(3)] Ask the first terms of the Magnus expansion \cite{Blanes2009} generated by $\he$ to match those of the desired evolution at the end of each period $T$, by enforcing equality through the amplitudes $\{\mathcal A_k,\mathcal B_k\}$ in the control functions.
\item[(4)] Interpolate solutions in different periods such as to obtain smooth control functions. This is straightforward in two-state problems for eCD schemes at first order in $T$.
\end{enumerate}
\begin{figure*}
\includegraphics[width=\textwidth]{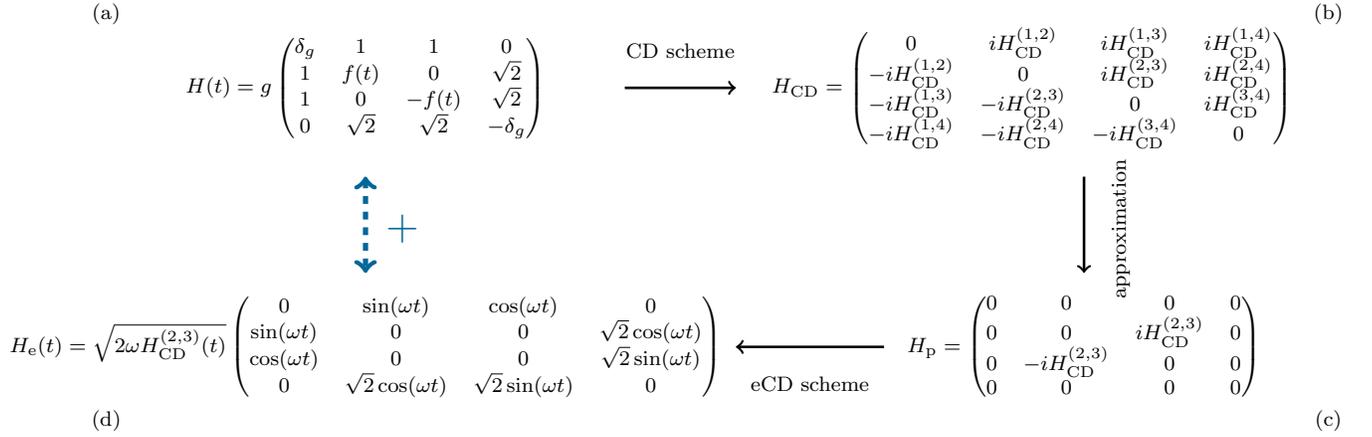}
\caption{Illustration of the construction of the eCD control method. After a reference adiabatic sweep function $f(t)$ is chosen (a), the corresponding CD Hamiltonian $\hcd$ is computed numerically, having the structure in panel (b). Next, neglecting transitions toward levels not involved in the exploited anticrossing, the corresponding CD corrections are set to zero (c), leading to the Hamiltonian $\hcdp$. Finally, the latter Hamiltonian is implemented effectively by introducing oscillating control functions in the qubit-resonator couplings (d), that produce the final correcting eCD Hamiltonian $\he$. The total Hamiltonian for the accelerated adiabatic driving is then $H(t) + \he(t)$.}
\label{fig:steps}
\end{figure*}
With the results of the previous section, let us compute the eCD Hamiltonian. One can verify that the matrix necessary to realize $\hcdp$, $\sigma_x^{(1)}\sigma_y^{(2)}-\sigma_y^{(1)}\sigma_x^{(2)}$, can be obtained as the commutator of the coupling matrices
$$2 i [C_1, C_2] = \sigma_y^{(1)}\sigma_x^{(2)}-\sigma_x^{(1)}\sigma_y^{(2)}. $$
This suggests choosing an eCD Hamiltonian of the form
$$ \he = c_1(t) C_1 + c_2(t) C_2 $$
with control functions $c_1(t)$ and $c_2(t)$ of the form in Eq. \eqref{eq:truncfour}. Fast oscillations in $c_1$ and $c_2$ are then tailored to effectively emulate the dynamics induced by $\hcdp$ of Eq. \eqref{eq:hcdp}. We make the simple choice
$$c_1(t) = \mathcal A \sqrt{\omega} \sin(\omega t);\quad c_2(t) = \mathcal B \sqrt{\omega} \cos(\omega t).$$
The proportionality of $c_1(t)$ and $c_2(t)$ to $\sqrt{\omega}$ is necessary for having the second term in the Magnus expansion to be of first order in $T$.
The exponent $\me$ of the Magnus expansion for the dynamics $\ue$ induced by $\he$, computed at the end of one period, is
\begin{equation}\label{eq:me}
\me(T) = - i \frac{\mathcal A \mathcal B}{4} T [\sigma_x^{(1)}\sigma_y^{(2)}-\sigma_y^{(1)}\sigma_x^{(2)}] + o(T^{3/2}),
\end{equation}
while the exponent $M_\text{p}$ for the Magnus expansion of $U_\text{p}$ induced by $H_\text{p}$ is, using Eq. \eqref{eq:hcdp},
\begin{align}
M_\text{p}(t) & = -i \int_0^T dt_1 \frac{\hcd^{(2,3)}}{2}(t_1)[\sigma_x^{(1)}\sigma_y^{(2)}-\sigma_y^{(1)}\sigma_x^{(2)}] . \nonumber \\
& = -i \frac{\hcd^{(2,3)}(T/2)}{2} T [\sigma_x^{(1)}\sigma_y^{(2)}-\sigma_y^{(1)}\sigma_x^{(2)}]+ o(T^3). \label{eq:mcd}
\end{align}
Let us remark that $\hcd^{(2,3)}$ is positive for all times if the initial adiabatic sweep function is $f_\text{PL}$; see Fig. \ref{fig:CDfield}. Moreover, for small $T$, it holds that $\hcd^{(2,3)}(T/2) = \hcd^{(2,3)}(t) + o(T^3)$. Therefore, a solution of the constraint equation $\me(T) \simeq M_\text{p}(T)$ to order $T$ can be obtained from Eqs. \eqref{eq:me} and \eqref{eq:mcd} as
\begin{equation}\label{eq:ecdampl}
\mathcal A(t) = \mathcal B(t) = \sqrt{2 \hcd^{(2,3)}(t)}.
\end{equation}
The complete eCD Hamiltonian then reads
\begin{equation}\label{eq:Ecdfield}
\he(t) = \sqrt{2 \omega \hcd^{(2,3)}(t)}\left[ \sin(\omega t) C_1 + \cos(\omega t) C_2\right].
\end{equation}
To clarify the whole procedure we illustrate in Fig. \ref{fig:steps} how the correcting Hamiltonians affect the matrix structure of the original Hamiltonian $H(t)$.

\section{Results}\label{sec:results}

The results given by the proposed quantum control method are reported in Figs. \ref{fig:ecd} and \ref{fig:ecd2}. In Fig. \ref{fig:ecd}, the eCD field is applied to assist an adiabatic sweep driven by the sweep function $f_\text{PL}$ of Eq. \eqref{eq:poly}, which proved in Sec. \ref{sec:sweeps} to be the best global adiabatic sweep; see Fig. \ref{fig:adiabatic}. In Fig. \ref{fig:ecd2}, the reference sweep is instead a local adiabatic one, $f_\text{tan}$ of Eq. \eqref{eq:qab}. We favored this one over $f_\text{RC}$ of Eq. \eqref{eq:rc} due to its stability with respect to changes in the total duration, see again Fig. \ref{fig:adiabatic}. 

Before discussing the results, let us explain how they are obtained and represented.
The parameters of the eCD corrections --- that is, amplitude and frequency --- are chosen such as to satisfy reasonable realizability constraints. The latter regard principally the strength of the time-dependent qubit-resonator coupling. The frequency must also be consistent with the RWA used in deriving the Hamiltonian of Eq. \eqref{eq:syst}; namely, it should not be as high as that of neglected terms and should not be resonant with unwanted transitions. For these reasons, in producing the colored solid lines in Figs. \ref{fig:ecd} and \ref{fig:ecd2}, two ceilings on the amplitude and frequency of the eCD field of Eq. \eqref{eq:Ecdfield} are imposed. For the amplitude, we impose that the maximal allowed value of $\sqrt{\omega} \mathcal A(t)$ of Eq. \eqref{eq:ecdampl} does not surpass a factor $k$ times the initial qubit-bus coupling $g$ --- that is, $\sqrt{\omega} \mathcal A(t) = k g$. Using Eq. \eqref{eq:Ecdfield}, this translates into an upper bound on $\omega$ for a given total duration $t_f$:
\begin{equation} \label{eq:om}
\omega = \frac{k^2  g^2 }{2 \max_t \hcd^{(2,3)}(t)}.
\end{equation}
Note that this expression is reminiscent of an adiabaticity condition; see Eq. \eqref{eq:adcond1} in Appendix \ref{sec:lad}. Together with Eq. \eqref{eq:om}, also the RWA must be taken into account. For this reason, even if Eq. \eqref{eq:om} is satisfied for a given $k$ and $t_f$, we impose a second ceiling on the frequency: The main (solid colored) curves are produced by requiring it not to surpass $\omega/2 \pi = 7$ GHz. This threshold is an arbitrary choice, made by considering that in this way the eCD controls are not resonant with any transition and that terms neglected by the RWA are of the order of $(\omega_r+\omega_{1,2})/2 \pi \simeq 14$ GHz with our working parameters. Nonetheless, in Figs. \ref{fig:ecd} and \ref{fig:ecd2}, one can easily extrapolate the results for different frequency ceilings by making use of the dotted, dashed, and dot-dashed black lines, which represent the eCD results for fixed $\omega/2\pi$ ranging from 4 to 8 GHz. More details about the RWA will be discussed in Sec. \ref{sec:param}. Summarizing, the solid colored eCD curves in Fig. \ref{fig:ecd} and \ref{fig:ecd2} are obtained by setting the eCD frequency $\omega$ to the largest value smaller or equal to $2\pi \times 7$ GHz that satisfies Eq. \eqref{eq:om}. The color palette then indicates the resulting maximal amplitude of the eCD control fields in units of $g$.

\subsection{Boundary cancellation method and eCD}

With the previous caveats in mind, one can see from Figs. \ref{fig:ecd} and \ref{fig:ecd2} that the eCD method helps to achieve an important speed up of the assisted adiabatic evolutions. Let us first discuss Fig. \ref{fig:ecd}, where the result of adding eCD corrections to a boundary cancellation method (BCM) is reported. The use of an eCD correcting field with maximal strength equal to $g$ [$k=1$ in Eq. \eqref{eq:om}], giving an overall qubit-bus coupling oscillating between 0 and 100 MHz, produces fidelities above 99.9\% with $\sim 30\%$ speed up with respect to the unassisted sweep. The total timescale in this case still remains above $2 \mu$s though, so stronger correcting fields are needed in order to access timescales of practical interest. If one allows the eCD field to reach maximal amplitudes of $2g$ [$k=2$ and 3 curves in Eq. \eqref{eq:om}] with frequency around $7$ GHz, the same fidelities can be reached, for instance, starting from $\sim 700-800$ ns. 

Let us note that, from Fig. \ref{fig:ecd}, two general behaviors can be observed. Initially, condition \eqref{eq:om} on $\omega$ is stricter than the ceiling at $7$ GHz: In this regime, the infidelity goes down rapidly with respect to the total time, even if following a polynomial scaling (linear in the logscale in Fig. \ref{fig:ecd}). When  condition \eqref{eq:om} starts to give frequencies larger than $7$ GHz, the $7$-GHz ceiling takes over, enforcing a slower, but exponential, decay of the infidelity. One can thus see that being able to reach higher ratios $k$ is mostly relevant for faster timescales.

\subsection{Local adiabatic driving and eCD}
\begin{figure}
\includegraphics[width=1.1\linewidth]{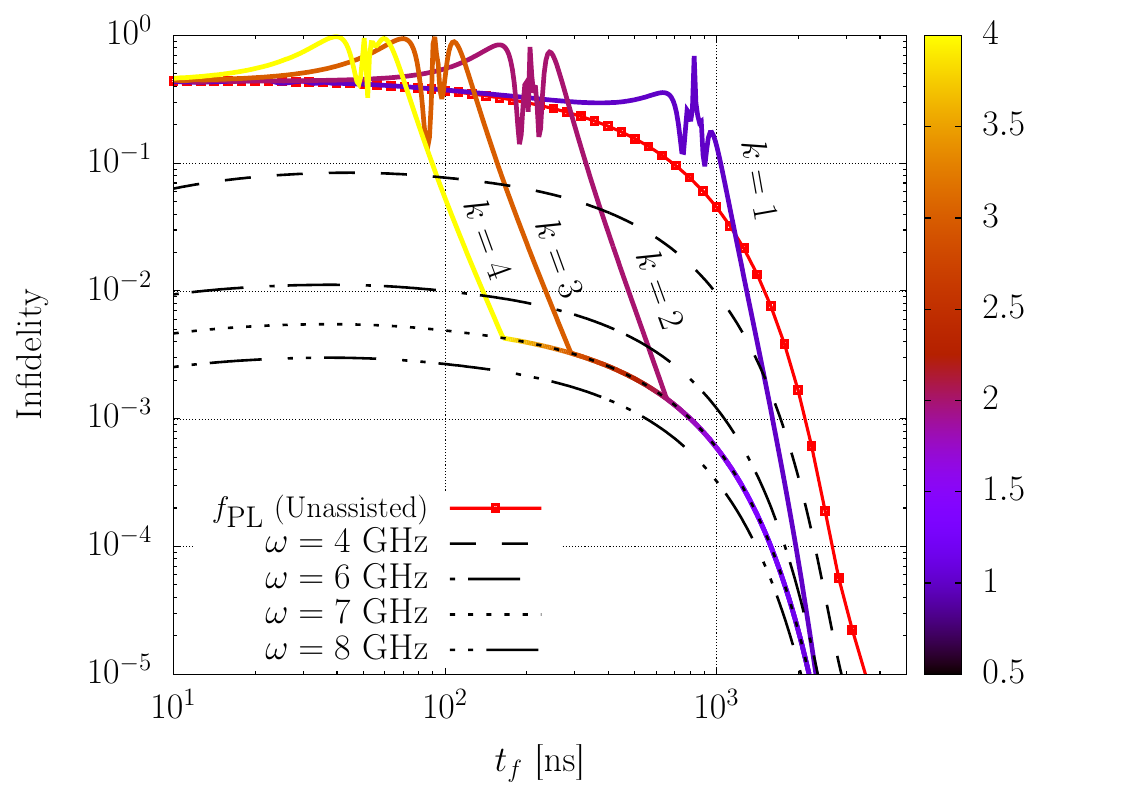}
\caption{Comparison between eCD and finite-time adiabatic driving, for a BCM sweep of the form of Eq. \eqref{eq:poly}, and for different ratios $k$ between the maximal amplitude of $\he$ and the initial qubit-bus coupling $g$; see Eqs. \eqref{eq:Ecdfield} and \eqref{eq:om}. The squares solid red line shows the results for the unassisted sweep, while the colored (shaded) solid lines correspond to different values of $k$, with an overall (angular) frequency ceiling at $2\pi \times 7$ GHz. The results for different frequency ceilings can be extrapolated using the dashed, dot-dashed, dotted and dot-dot-dashed curves, which represent the results produced by setting the eCD frequency $\omega/2\pi$ to 4, 5, 6, and 8 GHz, respectively. The color palette indicates the ratio between the eCD maximal amplitude and $g$.}
\label{fig:ecd}
\end{figure}

\begin{figure*}
\includegraphics[width=1.05\textwidth]{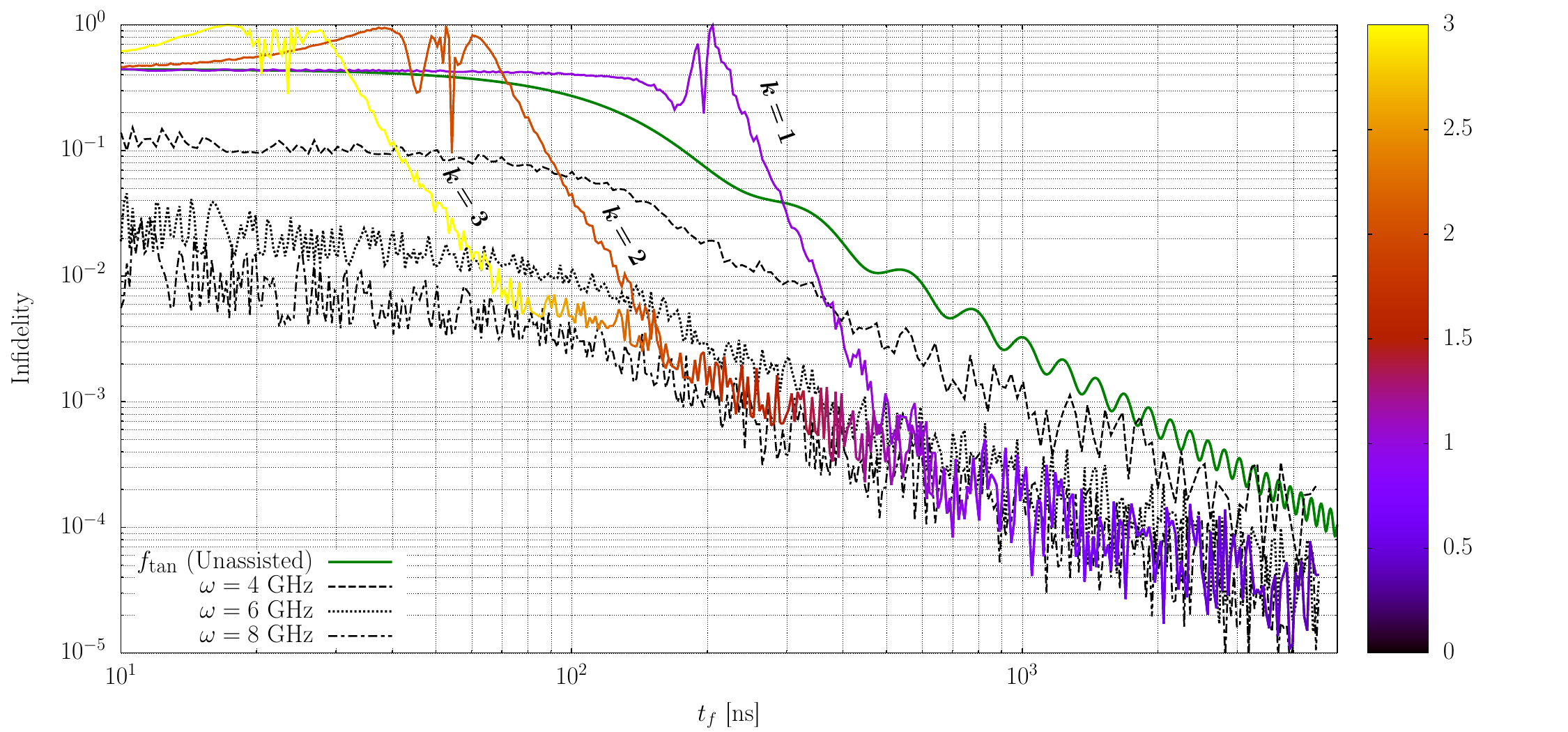}
\caption{Comparison between eCD and finite-time adiabatic driving, for a reference LAD of the form of Eq. \eqref{eq:qab}. Different ratios $k$ between the maximal amplitude of $\he$ and the initial qubit-bus coupling $g$ are considered; see Eqs. \eqref{eq:Ecdfield} and \eqref{eq:om}. The green solid line shows the results for the unassisted sweep. The colored (shaded) solid curves represent the eCD case for different $k$. The limit imposed by $k$ is further filtered by imposing an overall ceiling in the frequency of the eCD oscillations, set to $7$ GHz, for respecting reasonable realizability constraints and consistency with the RWA. The color palette indicates the ratio between eCD maximal amplitude and $g$. The broken black lines represent the infidelities produced by fixing the frequency at $4,$ 6, and 8 GHz and can be used for extrapolating the results given by different maximal frequency ceilings.}
\label{fig:ecd2}
\end{figure*}

Let us now turn to Fig. \ref{fig:ecd2}. In this case, the eCD corrections are applied to assist a LAD, namely $f_\text{tan}$ of Eq. \eqref{eq:qab}. One can immediately see that fidelities between 99\% and 99.99\% are achievable in shorter duration with respect to the BCM+eCD case of Fig. \ref{fig:ecd}. For higher fidelities, the BCM+eCD strategy gives better results instead. Using an eCD field with maximal amplitude equaling $g$ [$k=1$ in Eq. \eqref{eq:om}], fidelities above $99\%$ are accessible starting from less than $400$ ns, while $\sim 99.9\%$ fidelity is accessible in $\sim 500$ ns. Allowing the maximal amplitude to reach $2g$ and $3g$ gives instead more interesting results: Fidelities above $99\%$ are attained in $\sim 65$ and $\sim 130$ ns, respectively, while fidelities above $99.9\%$ are achievable in $\sim 400$ ns. Let us recall that all these results do not require eCD oscillation frequencies larger than $7$ GHz. As in the BCM+eCD case of the previous subsection, one can see that the $7$ GHz ceiling is dominant for large timescales, while admitting larger $k$ in Eq. \eqref{eq:om} determines how fast one can reach the ``$7$-GHz regime.'' Regimes relative to different maximal frequency ceilings, namely $\omega/2\pi=4,$ 5, 6, and $8$ GHz, can be seen from the black broken lines in Fig. \ref{fig:ecd2}: Increasing the maximal allowed frequency in general lowers the whole curve leading to higher fidelities for fixed duration.

This analysis shows that the LAD + eCD method is the most efficient protocol for the state transfer considered here in the timescales of practical interest, with respect to standard finite-time adiabatic driving, BCMs, and BCM + eCD. Comparing Figs. \ref{fig:ecd} and \ref{fig:ecd2}, we see that using the LAD + eCD strategy produces produces 99.9\% fidelities more than twenty times faster than a simple LZ sweep.
%
\subsection{Choosing the system parameters}\label{sec:param}

Let us now discuss how the results presented in the previous subsections depend on the specific experimental parameters used in the simulations (see the last paragraph of Sec. \ref{sec:specfram}). In general, for fixed detunings $\delta$ between qubits and resonator, raising the maximal values of the qubit-bus coupling $g$ would induce an improvement of all the methods discussed in this work in terms of the total duration needed for a desired infidelity, without changing much the comparison between different methods. This is so because, for the avoided crossing problem studied, raising the minimal gap naturally leads to smaller nonadiabatic errors and thus the global timescales for a given infidelity are shorter, provided $g$ does not become large enough to invalidate the RWA.

Changing the detunings of the qubits from the resonator can have a double effect. In fact, these have a role in (i) realizing the dispersive regime, which works when $g/\delta \ll 1$ and (ii) determining the strength of the resonator-mediated qubit-qubit coupling, which scales like $g^2/\delta_\text{res}$ where $\delta_{res}$ is the detuning of the two qubits from the resonator when in resonance with each other (see Appendix \ref{appendix:AC}). On the one hand, raising $\delta$ ulteriorly consolidates the dispersive regime, making the adiabatic elimination of states $\ket{0\uparrow \uparrow}$ and $\ket{2 \downarrow \downarrow}$ a better approximation. On the other hand, it would reduce the mediated coupling and thus amplify the timescales needed for adiabaticity. Therefore, for achieving the best performance with adiabatic methods in our setup one should find the smallest qubit-resonator detunings which give a satisfactory dispersive-regime approximation, while maximizing $g$ for having a larger avoided crossing. 

With the specific parameters used in this work, the RWA is a good approximation, in the sense that transitions outside the four-level subspace considered are negligible and conservation of the number of excitations does hold. This is also confirmed by numerical evidence. At the same time, the dispersive regime guarantees that the two levels involved in the avoided crossing exploited for the protocol are sufficiently decoupled from the other two. Nonetheless, counterrotating terms have still frequencies less than ten times greater than those of rotating ones [$|\omega_k-\omega_r|/2\pi \simeq 2$ GHz, $(\omega_k+\omega_r)/2\pi \simeq 14$ GHz]. As a consequence, considering the full spectrum has an appreciable effect in renormalizing the width of the avoided crossing. More specifically, when $\omega_r < \omega_k$ the gap tends to reduce, while it becomes larger if $\omega_r>\omega_k$ as in our case. This is explained by a perturbative calculation in Appendix \ref{appendix:AC}. Importantly, taking into account the full spectrum can thus even cause an improvement of all the methods described in this work, thanks to the enlargement of the anticrossing. These features have also been seen in numerical simulations.

Let us now discuss the parameter $g f_0$, which quantifies how far the two qubits are from being in resonance at the beginning of the protocol. Smaller values are acceptable unless one starts very close to the avoided crossing, and as the value is decreased all the methods would tend to give the same results, as the whole driving gets closer to a local LZ linear regime. Larger values would make the LAD methods work even better with respect to BCMs. Indeed, by further accelerating in the beginning where nonadiabatic effects are almost negligible, the LADs would need a similar total time for achieving a given fidelity as with the parameters used here.

\section{Robustness}\label{sec:stability}

\begin{figure}
\includegraphics[width=\linewidth]{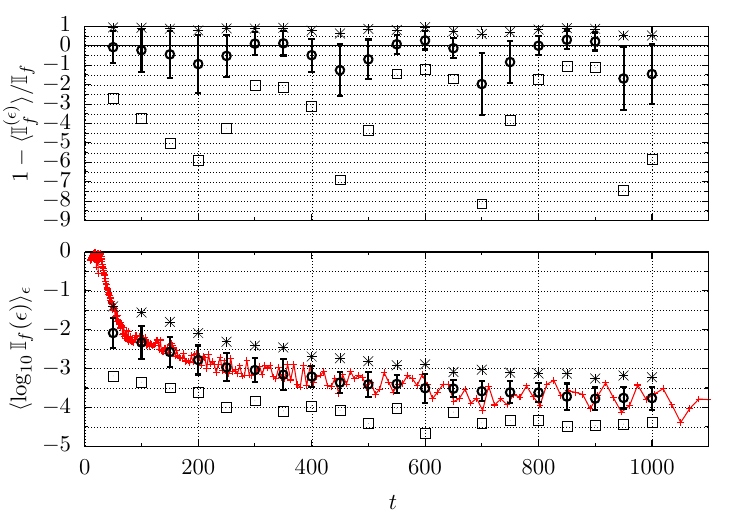}
\caption{In the top panel, the average relative error on the final infidelity is shown (empty circles), for different total duration of the protocol, when random imperfections within $5\%$ in relative value are simultaneously present in frequency, amplitude and relative phase [see Eqs. \eqref{eq:errpar} and \eqref{eq:infeps}] of the eCD control functions. The scheme is the LAD + eCD driving described in Sec. \ref{sec:results} with unperturbed eCD frequency fixed at $\omega/2\pi=7$ GHz. 
In the lower panel, the average logarithmic infidelity for the same data is represented (empty circles), with the solid red line reporting the unperturbed results. In both panels, the error bars extend to one standard deviation, while the asterisk and squared points indicate maximal and minimal values, respectively.}
\label{fig:allerrs}
\end{figure}
In this section, we discuss the robustness of the proposed quantum control method against possible sources of errors. In particular, we first study the effect of potential static shifts in the relevant parameters of the protocol and, following, we discuss damping and decoherence effects arising from the interaction of the qubit-resonator system with the environment.
This analysis will be done for the case of the LAD + eCD sweep studied in Sec. \ref{sec:results}, for $\omega/2\pi = 7$ GHz, as represented in Fig. \ref{fig:ecd2}, since it has shown to give the best results in terms of fidelity for a given total protocol time.


We start by analyzing numerically how random static shifts in the frequency, amplitude and relative phase of the eCD control fields affect the final fidelity. With reference to the eCD Hamiltonian of Eqs. \eqref{eq:ecdampl} and \eqref{eq:Ecdfield}, this is done by adding a small perturbation to each one of the corresponding parameters, namely
\begin{align}
    \omega & \to \omega [1+\epsilon_\omega], \nonumber \\
    \cos(\omega t) & \to \cos (\omega t +  \pi \epsilon_\phi),\label{eq:errpar} \\
    \mathcal A(t) & \to\mathcal A(t)[1+\epsilon_\mathcal{A}] \nonumber .
\end{align}
The perturbations $\bm{\epsilon}=\{\epsilon_\omega$, $\epsilon_\phi$, $\epsilon_\mathcal{A}\}$ are taken to be random variables uniformly distributed in an interval $[-\epsilon_\text{max},+\epsilon_\text{max}]$. We choose specifically $\epsilon_\text{max}=5\%$.
The effect of the imperfections on the system evolution can be quantified by studying the average final infidelity $\av$, for a fixed total duration $t_f$, where the average $\langle \cdot \rangle_{\bm{\epsilon}}$ is taken over $N_{\bm{\epsilon}}$ realizations of the variables $\bm{\epsilon}$,
\begin{equation}\label{eq:infeps}
\av = \frac{1}{N_{\bm{\epsilon}}} \sum_{\bm{\epsilon}} \If(\bm{\epsilon}),
\end{equation}
and $\If(\bm{\epsilon})$ is the final infidelity in the presence of error $\epsilon$.
The quantity we monitor is then the relative deviation between the average final infidelity $\av$ and the final infidelity $\If^{(0)}$ in absence of errors $[\bm{\epsilon}=\bm{0}]$:
$$1-\frac{\av}{\If^{(0)}} .$$

In Figure \ref{fig:allerrs}, this quantity is plotted for changing total duration and for errors present in all parameters, together with the maximal and minimal value and the standard deviation
$$\sigma(\If)=\sqrt{\langle \If^2(\bm{\epsilon})\rangle_\epsilon-\langle \If(\bm{\epsilon})\rangle_\epsilon^2}.$$
Both max/min value (asterisks and squares, respectively) and $\sigma$ (error bars) are represented so that one can estimate the worst-case scenario while figuring out what to expect on average.

From the top panel of Fig. \ref{fig:allerrs}, one can see that the average value is typically negative: As naively expected, unpredicted variations typically hinder the performance leading to higher infidelities for fixed duration. Nonetheless, the error bars, together with the maximal values, witness that there are some cases in which the fidelity increases in the presence of parameter biases. Considering the procedure described in Sec \ref{sec:ecd} for developing the eCD field, this can happen when small errors lead to an accidental improvement of the approximate equality of the first Magnus terms~\cite{PetWimb2019}. Moreover, since the eCD dynamics oscillates quickly around the desired one, small biases can bring it closer to the target in a given total time.

Worst-case values present infidelities up to $\sim$6 times greater than the unperturbed case, while the best case can improve up to 100\%. The standard deviation is always less than $100\%$. Remarkably, this shows that, although the exact value of the infidelity is quite unstable, the order of magnitude of the final fidelity is rather robust: Errors within $5\%$ in system parameters never produce a jump of the order of magnitude.

To better inspect this feature, we study the average value of the logarithm of the final infidelity, 
$$\langle \log_{10} \If \rangle_\epsilon.$$
This is shown in the lower panel of Fig. \ref{fig:allerrs} for different values of the total duration, compared with the unperturbed values. The error bars indicate again one standard deviation, $\sigma(X) = \sqrt{\langle X^2\rangle_\epsilon-\langle X \rangle_\epsilon^2}$ with $X = \log_{10}[\If(\bm{\epsilon})]$. The average values are very close to the unperturbed curve, and $\sigma$ shows that, as previously observed, the difference between perturbed and unperturbed values is always smaller than one order of magnitude. This gives a good evidence of the robustness of the method: If one is interested in maintaining a certain fidelity threshold of the state transfer, small parameter biases (within $5\%$ here) need not be taken care of. For instance, from Fig. \ref{fig:allerrs}, we see that fidelities above $99.9\%$ are achievable in $>400$ ns even in the presence of imperfections.

\begin{figure}
\includegraphics[width=\linewidth]{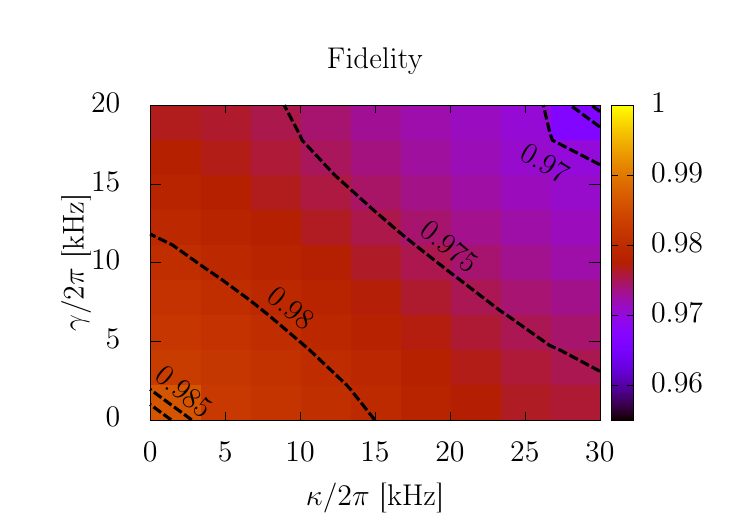}
\caption{Final fidelity obtained with the proposed control protocol in the presence of damping and decoherence for different values of the decay rates. This is obtained by numerical solution of the master Eq. \eqref{eq:lindblad} for a total duration $t_f=100$ ns, assuming equal rates for the two qubits $\gamma_r^{(1)}=\gamma_r^{(2)}\equiv \gamma$ and $\gamma_\phi^{(1)}=\gamma_\phi^{(2)}=\gamma/2$. For typical experimental values of $\kappa$ and $\gamma$, the fidelity remains around $98\%$, and even letting the rates reach more pessimistic values, it does not drop below $97\%$, proving a good robustness of the control method.}
\label{fig:diss}
\end{figure}

Let us now turn to a discussion of dissipation and decoherence effects. To this end, we incorporate into the description of the evolution of the system the spontaneous emission and decoherence of the two qubits and photon leakage from the resonator. This is done by describing the evolution of the density matrix $\rho(t)$ of the system through the following master equation in Lindblad form \cite{Blais2007}:
\begin{multline} \label{eq:lindblad}
\frac{\partial \rho}{\partial t} = -i[H(t),\rho] + \kappa D[a] \rho +\\
 \sum_{k=1}^2 \gamma_r^{(k)} D[\sigma_-^{(k)}] \rho + \sum_{k=1}^{2} \gamma_\phi^{(k)} D[\sigma_z^{(k)}] \rho,
 \end{multline}
where $\kappa$ is the damping rate for the resonator, $\gamma_r^{(k)}$ and $\gamma_\phi^{(k)}$ are the relaxation and dephasing rates, respectively, for qubit $k$ and $\mathcal D[X] = X \rho X^\dagger - \frac{1}{2} \{X^\dagger X, \rho\} $ is the Lindblad dissipator. The Hamiltonian $H(t)$ in Eq. \eqref{eq:lindblad} is the full Hamiltonian of the controlled system in the RWA, including the eCD correction.
Figure \ref{fig:diss} shows the final fidelity of the protocol for a representative value of the total duration $t_f=100$ ns and for realistic parametric ranges of the rates. This is obtained by numerically solving the master Eq. \eqref{eq:lindblad} with three photonic states of the resonator. For simplicity, equal rates for the two qubits are assumed, with $\gamma_r^{(1)}=\gamma_r^{(2)}\equiv \gamma$ and $\gamma_\phi^{(1)}=\gamma_\phi^{(2)}=\gamma/2$.
For typical values $\kappa/2\pi= 5$ kHz and $\gamma/2\pi = 5$ kHz, the fidelity is well above $98\%$. Considering rates above $10-15$ kHz makes the fidelity drop below $98\%$, but even for large values it remains above $97\%$. These results show that the proposed control method is rather reliable even in the presence of damping and decoherence effects. 

\section{Conclusion}\label{sec:conclusion}

An assisted adiabatic passage has been theoretically proposed as a strategy for entangling two qubits in a cQED architecture, with the qubits dispersively coupled to a coplanar waveguide resonator. The method works by slowly driving the qubits into resonance with each other, so that the adiabatic path connects a separable initial eigenstate to a final Bell entangled state (eigenstate of the dispersive flip-flop interaction). The adiabatic passage is accelerated by adding fast oscillations in the qubit-resonator couplings which counteract nonadiabatic transitions, as recently proposed in Ref. \cite{petiziol2018}.

The method achieves both of the desired goals, high fidelities and a short gate time, while requiring only modest extra control resources. Furthermore, it inherits the robustness against unpredicted shifts in control parameters from traditional adiabatic methods, and proves to be stable with respect to damping and decoherence effects. As such, this is expected to be of great practical use in entangling qubits in a standard cQED infrastructure, or for other systems well described by a Jaynes-Cummings Hamiltonian.

It has been shown that best results are obtained by oscillating around a local adiabatic driving, that is, an adiabatic driving whose sweep rate is adapted so to satisfy a local adiabatic condition in each instant of time. Moreover, not only does the protocol work within the rotating-wave approximation, despite requiring fast oscillations, but the presence of the full spectrum can lead to better performances. This happens since, with the parameters used, including counter-rotating terms induces a Bloch-Siegert shift which can enlarge the anticrossing, thus globally reducing the timescales needed for adiabaticity.


\acknowledgments
We gratefully acknowledge financial support within the Imperial College European Partners programme.
This research benefits from the HPC (High Performance Computing) facility of the University of Parma, Italy.
\appendix
\section{Matrices} \label{sec:matrices}

In this appendix we introduce the following matrices, which are referred to in the main text in Sec. \ref{sec:specfram} after Eq. \eqref{eq:syst}, when the Jaynes-Cummings Hamiltonian of Eq. \eqref{eq:hjc} is reduced to the four-by-four matrix of Eq. \eqref{eq:ham} by considering the subspace of two excitations:
$$ D_0 = \begin{pmatrix}
1 & 0 & 0 & 0 \\
0 & 0 & 0 & 0 \\
0 & 0 & 0 & 0 \\
0 & 0 & 0 & -1
\end{pmatrix}, \quad D_1 = \begin{pmatrix}
0 & 0 & 0 & 0 \\
0 & 1 & 0 & 0 \\
0 & 0 & -1 & 0 \\
0 & 0 & 0 & 0
\end{pmatrix},$$
$$ C_1 = \begin{pmatrix}
0 & 1 & 0 & 0 \\
1 & 0 & 0 & 0 \\
0 & 0 & 0 & \sqrt{2} \\
0 & 0 & \sqrt{2} & 0
\end{pmatrix} ,\quad C_2 = \begin{pmatrix}
0 & 0 & 1 & 0 \\
0 & 0 & 0 & \sqrt{2} \\
1 & 0 & 0 & 0 \\
0 & \sqrt{2} & 0 & 0
\end{pmatrix}.$$

\section{Local adiabatic driving} \label{sec:lad}

In this appendix, we recall the main ideas behind the local adiabatic driving (LAD) strategies, and we derive the sweep functions used in the main text, Eqs. \eqref{eq:rc} and \eqref{eq:qab}.
For relevance to our purpose, we treat directly the case of a two-state avoided-crossing problem, specifically with Hamiltonian ($\hbar=1$)
\begin{equation}\label{eq:hama}
H(t) = g f(t) \sigma_z + x_0 \sigma_x,
\end{equation} 
with $g>0$, $f(t)\ge 0$ dimensionless, and $\partial_t f(t)\le 0$. This writing of the two-level Hamiltonian is helpful for mapping its parameters to those used in the main text for the Hamiltonian of Eq. \eqref{eq:ham} and the sweep functions of Sec. \ref{sec:sweeps}. Namely, $g$ and $f(t)$ can be directly mapped to $g$ and $f(t)$ of Eqs. \eqref{eq:ham} and \eqref{eq:sweep0}, respectively, while for the parameter $g_0$ introduced in Sec. \ref{sec:specfram} it holds $g_0=|x_0|$.

Let us suppose that the system state $\ket{\psi(t)}$ is initially in the instantaneous ground state $\ket{gs(0)}$ of the Hamiltonian $H(0)$, $\ket{\psi(0)}=\ket{gs(0)}$. The adiabatic theorem \cite{messiah1961qm} states that, if the adiabatic condition
\begin{equation}\label{eq:adcond1}
\frac{\max_t \lvert \bra{e(t)} \partial_t H \ket{gs(t)}\lvert}{\min_t \lvert E_{gs}(t)-E_{e}(t)\lvert^2} < \varepsilon
\end{equation}
is satisfied, with $\ket{e(t)}$ being the excited instantaneous eigenvector at time $t$ and $E_e(t), E_{gs}(t)$ being the instantaneous eigenenergies, then the final fidelity is 
$$\mathbb{F}(t_f) =\lvert \braket{\psi(t_f)\lvert gs(t_f)}\lvert^2>1-\varepsilon^2.$$ 
The idea which was put forward in Ref. \cite{Roland2002} is that the above reasoning can be applied to every arbitrarily small time interval $T$ of the total evolution. Then, one may adapt the speed of the evolution in such a way that the condition \eqref{eq:adcond1} is satisfied for the interval of width $T$. In the limit of vanishing $T$, this produces a local form of the adiabatic condition \cite{Roland2002,Amin2009}:
\begin{equation}\label{eq:adcond2}
\left\lvert\frac{df}{dt} \right\lvert\frac{\lvert \bra{e(f)} \partial_f H\ket{gs(f)}\lvert}{ \left[E_{gs}(f)-E_e(f)\right]^2} < \varepsilon.
\end{equation}
Then, turning the inequality into an equality, one can solve the resulting differential equation to find $t(f)$ such that the condition is satisfied. This can then be inverted in some cases to find the sweep function $f(t)$. Here we discuss the use of two different adiabatic conditions to obtain the corresponding sweep functions $f(t)$.

\subsection{Roland-Cerf function}

With the Hamiltonian of Eq. \eqref{eq:hama}, the instantaneous gap is
$$\Delta E[f(t)] = 2 g \sqrt{f(t)^2+(x_0/g)^2} $$
and the nonadiabatic coupling gives
$$ \lvert \bra{e[f(t)]} \partial_f H[f(t)] \ket{gs[f(t)]} \lvert=  \frac{|x_0|}{\sqrt{f^2(t)+(x_0/g)^2}}.$$
The local adiabatic condition \eqref{eq:adcond2}, assuming $f(t)$ to be monotonically non-increasing $df/dt\le 0$ with $f(0)=f_0>0$, produces the relation
$$\frac{dt}{df} = -\frac{|x_0|}{4 \epsilon g^2 [f^2(t)+(x_0/g)^2]^{3/2}}. $$
Once integrated, taking into account that $f(t_f)=0$, one obtains that the total time is
$$ t_f = \frac{f_0}{4 \epsilon |x_0| \sqrt{f_0^2+(x_0/g)^2}}. $$
The sweep function, introducing the rescaled time $s=t/t_f$, is
\begin{equation*}
f(s) = \frac{|x_0 |f_0(1-s)}{g \sqrt{(x_0/g)^2+f_0^2 s(2-s)}}.
\end{equation*}

\subsection{Quantum adiabatic brachistochrone}

In Ref. \cite{Rezakhani2009}, it is pointed out that more accurate adiabatic conditions typically involve powers of the norm of the derivative of the Hamiltonian, rather than single matrix elements. For this reason and to be able to formulate the problem in a geometric context, Ref. \cite{Rezakhani2009} proposes the condition
$$\left\lvert \frac{df}{dt} \right\lvert \frac{\lVert \partial_f H[f(t)] \lVert}{ \Delta E[f(t)]^2} < \varepsilon,$$
where $\lVert A \lVert = \sqrt{\text{tr}[A A^\dagger]}$ is the Hilbert-Schmidt norm.
Acting as above to find $f(t)$, the latter choice, for $H(t)$ of Eq. \eqref{eq:hama}, gives the differential equation
$$ \frac{dt}{df} =- \frac{\sqrt{2}}{4 \epsilon g [f^2(t)+(x_0/g)^2]} .$$
The solution is
$$f(s) = \frac{x_0}{g} \tan \left[ \alpha  (1-s)\right] $$
for $s=t/t_f$ and $\alpha= \arctan(g f_0/x_0)$.
The total time is
$$t_f = \frac{\alpha}{2\sqrt{2} \epsilon x_0}.$$
Let us recall that $f_0>0$ and $g>0$ so $t_f$ is actually positive for all values of $x_0$.

\section{Anticrossing formation} \label{appendix:AC}

The purpose of this appendix is to discuss the formation of the anticrossing which is exploited for the adiabatic methods described in the main text. The qubit-qubit coupling mediated by the resonator, responsible for this phenomenon, is derived perturbatively in the RWA first, following Ref. \cite{Blais2004}, and without the RWA afterward. 
Let us consider the full Rabi Hamiltonian 
\begin{multline} H_\text{R} = \omega_r a^\dagger a + \sum_k \frac{\omega_k}{2} \sz{k}+ \sum_k g (a \splus{k}+a^\dagger \smin{k})\\
+\sum_k g (a^\dagger \splus{k}+a \smin{k}).\end{multline}
Introducing the unitary 
$$U_1 =\exp\left[ -\sum_k \frac{g}{\delta_k} (a^\dagger \smin{k}-a \splus{k})\right] ,$$
with $\delta_k = \omega_k-\omega_r$, the Hamiltonian $U_1 H_\text{R} U_1^\dagger$, to second order in $g/\delta_k$, becomes
\begin{widetext}
\begin{align}
U_1 H_\text{R} U_1^\dagger &=  \omega_r a^\dagger a + \sum_k \left\{ \frac{\omega_k}{2} +\frac{g^2}{2\delta_k} (2 a^\dagger a+1) \right\} \sz{k}+ \frac{1}{2}\sum_{k,n} \frac{g^2}{\delta_k} \left[ \smin{n}\splus{k}+\splus{n}\smin{k}\right]\nonumber\\
& +\sum_k g (a^\dagger \splus{k}+a \smin{k}) +\sum_k \frac{g^2}{\delta_k} [a^2+(a^\dagger)^2]\sz{k} + \sum_{k,n} \frac{g^2}{\delta_k} \left[ \smin{n}\smin{k}+\splus{n}\splus{k}\right]. \label{ac:rwa}
\end{align} 
\end{widetext}
The first line in Eq. \eqref{ac:rwa} would be the result in the RWA, while the second line constitutes the byproduct of keeping counter-rotating terms. The last term in the first line describes the production of the anticrossing, which is exploited in the adiabatic methods described in the main text. When the qubits are in resonance with each other, $\delta_1=\delta_2\equiv \delta$, we thus find a coupling $ \frac{g^2}{\delta}$ (to second order), which for our parameters gives the value $\sim-7.1$ MHz, i.e., a minimal gap of $\sim 14.2$ MHz. In order to understand the contribution to the anticrossing of the counterrotating terms, let us make a second basis transformation determined by the unitary matrix
$$
U_2 = \exp\left[ -\sum_k \frac{g}{\Delta_k} (a \smin{k}-a^\dagger \splus{k})\right],
$$
where $\Delta_k = \omega_k+\omega_r$.
This is constructed so to eliminate the remaining (counterrotating) $\propto g$ term in the second line of Eq. \eqref{ac:rwa}. Again to second order, one has:
\begin{widetext}
\begin{align*}
U_2 U_1 H_\text{R} U_1^\dagger U_2^\dagger &=  \omega_r a^\dagger a +\sum_k \left\{ \frac{\omega_k}{2} +\frac{1}{2}\left(\frac{g^2}{\delta_k}+\frac{g^2}{\Delta_k}\right) (2 a^\dagger a+1) \right\} \sz{k}+ \frac{1}{2}\sum_{k,n} \left(\frac{g^2}{\delta_k}-\frac{g^2}{\Delta_k} \right)\left[ \smin{n}\splus{k}+\splus{n}\smin{k}\right]\\
&  +\sum_k \frac{g^2}{\delta_k} [a^2+(a^\dagger)^2]\sz{k} + \sum_{k,n} \frac{g^2}{\delta_k} \left[ \smin{n}\smin{k}+\splus{n}\splus{k}\right].
\end{align*} 
\end{widetext}

From the third term in the first line, we now see that the coupling is renormalized, due to the inclusion of counter-rotating terms, to $g^2\left(\frac{1}{\delta} -\frac{1}{\Delta} \right)$ (for $\Delta_1=\Delta_2\equiv\Delta$). Depending on whether the resonator has a higher or lower transition frequency with respect to the two qubits, the new shift can enhance or reduce the strength (absolute value) of the coupling. For our protocol, the first situation is the ideal one: with our parameters, the renormalized coupling is $\sim -8.2$ MHz. Keeping the same $\omega_k$ and $|\delta_k|$ for the second situation would give $\omega_r =3.8$ GHz and the coupling would reduce in strength to $\sim-5.5$ MHz. 


%

\end{document}